\documentclass[a4paper,fleqn]{cas-sc}

\usepackage[authoryear]{natbib}
\usepackage{float}
\usepackage{lineno}
\def\tsc#1{\csdef{#1}{\textsc{\lowercase{#1}}\xspace}}
\tsc{WGM}
\tsc{QE}
\begin{document}
\let\WriteBookmarks\relax
\def\floatpagepagefraction{1}
\def\textpagefraction{.001}

\shorttitle{The fall of asteroid 2024~XA$_1$ and the location of possible meteorites}    

\shortauthors{Gianotto et al.}  

\title[mode = title]{The fall of asteroid 2024~XA$_1$ and the location of possible meteorites}


%\tnotetext[1]{<tnote text>} 


\author[1,2]{Francesco Gianotto}[orcid=0009-0005-1927-6757]
\author[3]{Albino Carbognani}[orcid=0000-0002-0737-7068]
\author[1,4]{Marco Fenucci}[orcid=0000-0002-7058-0413]
\ead{marco.fenucci@ext.esa.int}
\author[1,2]{Maxime Devog\`{e}le}[orcid=0000-0002-6509-6360]
\author[5,6]{Pablo Ramirez-Moreta}[orcid=0000-0001-5874-6057]
\author[1,2]{Marco Micheli}[orcid=0000-0001-7895-8209]
\author[7]{Raffaele Salerno}
\author[8,9]{Toni Santana-Ros}[orcid=0000-0002-0143-9440]
\author[10]{Juan Luis Cano}[orcid=0000-0002-2005-4255]
\author[1]{Luca Conversi}[orcid=0000-0002-6710-8476]
\author[1]{Charlie Drury}[orcid=0009-0003-0296-3310]
\author[1,2]{Laura Faggioli}[orcid=0000-0002-5447-432X]          
\author[1,2]{Dora F\"{o}hring}[orcid=0000-0001-9259-2688]
\author[10,11]{Reiner Kresken}
\author[1,12]{Selina Machnitzky}
\author[1]{Richard Moissl}
\author[5,6]{Francisco Oca\~{n}a}[orcid=0000-0002-9836-3285]
\author[1,4]{Dario Oliviero}
\author[1,13]{Eduardo Alonso-Peleato}
\author[1,14]{Margherita Revellino}
\author[15,16]{Regina Rudawska}


\affiliation[1]{organization={ESA ESRIN / PDO / NEO Coordination Centre},
            addressline={Largo Galileo Galilei, 1}, 
            city={Frascati (RM)},
            citysep={}, % Uncomment if no comma needed between city and postcode
            postcode={00044}, 
            country={Italy}}
            
\affiliation[2]{organization={Starion Italia},
            addressline={Via di Grotte Portella, 28}, 
            city={Frascati (RM)},
            citysep={}, % Uncomment if no comma needed between city and postcode
            postcode={00044}, 
            country={Italy}}
\affiliation[3]{organization={INAF - Osservatorio di Astrofisica e Scienza dello Spazio},
            addressline={Via Gobetti, 93/3}, 
            city={Bologna},
            citysep={}, % Uncomment if no comma needed between city and postcode
            postcode={40129}, 
            country={Italy}}
\affiliation[4]{organization={Deimos Italia s.r.l.},
            addressline={Via Alcide De Gasperi, 24}, 
            city={San Pietro Mosezzo (NO)},
            citysep={}, % Uncomment if no comma needed between city and postcode
            postcode={28060}, 
            country={Italy}}
\affiliation[5]{organization={ESA ESAC / PDO Villafranca del Castillo},
            addressline={Bajo del Castillo, s/n}, 
            city={Madrid},
            citysep={}, % Uncomment if no comma needed between city and postcode
            postcode={28692}, 
            country={Spain}}
\affiliation[6]{organization={Deimos Space S.L.U.},
            addressline={Ronda de Poniente, 19}, 
            city={Tres Cantos Madrid},
            citysep={}, % Uncomment if no comma needed between city and postcode
            postcode={28760}, 
            country={Spain}}
\affiliation[7]{organization={Meteo Expert},
            addressline={Via G. Marconi, 27}, 
            city={Milano},
            citysep={}, % Uncomment if no comma needed between city and postcode
            postcode={20054}, 
            country={Italy}}
\affiliation[8]{organization={Departamento de Fisica, Ingeniería de Sistemas y Teoría de la Señal, Universidad de Alicante},
            addressline={Carr. de San Vicente del Raspeig, s/n}, 
            city={San Vicente del Raspeig, Alicante},
            citysep={}, % Uncomment if no comma needed between city and postcode
            postcode={03690}, 
            country={Spain}}
\affiliation[9]{organization={Institut de Ci\`{e}ncies del Cosmos (ICCUB), Universitat de Barcelona (IEEC-UB)},
            addressline={ Carrer de Mart\'{\i} i Franqu\`{e}s, 1}, 
            city={Barcelona},
            citysep={}, % Uncomment if no comma needed between city and postcode
            postcode={08028}, 
            country={Spain}}
\affiliation[10]{organization={ESA ESOC / PDO},
            addressline={ Robert-Bosch-Straße 5}, 
            city={Darmstadt},
            citysep={}, % Uncomment if no comma needed between city and postcode
            postcode={64293}, 
            country={ Germany}}
\affiliation[11]{organization={CGI Deutschland B.V. \& Ko. KG},
            addressline={Rheinstrasse 95}, 
            city={Darmstadt},
            citysep={}, % Uncomment if no comma needed between city and postcode
            postcode={64295}, 
            country={Germany}}
\affiliation[12]{organization={University of Freiburg},
            addressline={Fahnenbergplatz}, 
            city={Freiburg im Breisgau},
            citysep={}, % Uncomment if no comma needed between city and postcode
            postcode={79085}, 
            country={Germany}}
\affiliation[13]{organization={Alia Space System Srl},
            addressline={Via San Giuseppe Calasanzio 15}, 
            city={Frascati (RM)},
            citysep={}, % Uncomment if no comma needed between city and postcode
            postcode={00044}, 
            country={ Italy}}
\affiliation[14]{organization={Technische Universiteit Delft},
            addressline={Mekelweg 5}, 
            city={CD Delft},
            citysep={}, % Uncomment if no comma needed between city and postcode
            postcode={2628}, 
            country={ The Netherlands}}
\affiliation[15]{organization={ESA ESTEC / PDO},
            addressline={ Keplerlaan 1}, 
            city={AZ Noordwijk},
            citysep={}, % Uncomment if no comma needed between city and postcode
            postcode={2201}, 
            country={ The Netherlands}}
\affiliation[16]{organization={Starion Netherlands},
            addressline={Schuttersveld 2}, 
            city={ZA, Leiden,},
            citysep={}, % Uncomment if no comma needed between city and postcode
            postcode={2316}, 
            country={ The Netherlands}}

\affiliation[1]{organization={INAF - Osservatorio di Astrofisica e Scienza dello Spazio},
            addressline={Via Gobetti 93/3 }, 
            city={Bologna},
            citysep={}, % Uncomment if no comma needed between city and postcode
            postcode={40129}, 
            country={Italy}}

\cortext[1]{Corresponding author:}



\begin{abstract}
     Asteroid 2024~XA$_1$ was discovered on 3 December 2024 at 05:54 UTC by the Bok telescope in Kitt Peak, Arizona, and impacted Earth about 10 hours later over a remote area of the Sakha Republic (Russia). The estimated size of the object was about one meter, and the atmospheric entry produced a bright fireball that was captured by a webcam and several eyewitnesses. The first impact alert was issued at 07:50 UTC by the Meerkat Asteroid Guard of the European Space Agency, which triggered subsequent follow-up observations that confirmed both the object to be real and the occurrence of the impact with Earth. Here we present the operations and results from the NEO Coordination Centre (NEOCC) upon the impact event. Because the entry likely dropped meteorites on the ground, we also estimate the possible strewn fields for future meteorite search campaigns.
\end{abstract}

%\item Asteroid 2024~XA$_1$ was discovered on 3 December 2024, and it impacted Earth 10 hours later.
%\item Impact alarms from ESA Meerkat triggered the follow-up of the object.
%\item We determined the orbit and impact point of 2024~XA$_1$.
%\item We give a possible strewn fields for the fall of asteroid 2024~XA$_1$.

% Keywords
% Each keyword is separated by \sep
\begin{keywords}
 minor planets, asteroids: individual: 2024~XA$_1$ \sep astrometry \sep Meteorites, meteors, meteoroids
\end{keywords}

\maketitle

\section{Introduction}
\label{s:intro}

%%%%%%%%%%%%%%%%
Several surveys dedicated to the discovery of new near-Earth asteroids (NEAs)
$-$ such as the Catalina Sky Survey \citep{fuls-etal_2023}, the Panoramic Survey Telescope and Rapid Response System \citep[Pan-STARRS,][]{denneau-etal_2013}, and the Asteroid Terrestrial-impact Last Alert System \citep[ATLAS,][]{tonry-etal_2018} $-$ became operational in the last 20 years, and the number of known
NEAs grew significantly. Thanks to these facilities, asteroid 2008~TC$_3$ was
discovered 20 hours before impacting over Sudan \citep{jenniskens-etal_2009}, and it was the first
NEA discovered before impact ever. Because of the short time between discovery
and impact, 2008~TC$_3$ has been called also an \textit{imminent impactor}.
Since then, the discovery of imminent impactors is an increasingly frequent
event, as reported in Tab.~\ref{tab:small_impact}. Other ground based surveys
are currently under construction, such as the Vera Rubin Observatory \citep{ivezic-etal_2019} and
the Flyeye telescope \citep{conversi-etal_2021, fohring-etal_2024} by the European Space Agency (ESA), while the
space based telescopes NEO Surveyor Mission \citep[NEOSM,][]{mainzer-etal_2023} by NASA, and the NEO
Mission in the Infrared \citep[NEOMIR,][]{conversi-etal_2023} by ESA, are foreseen in the next
decade. Thus, the rate of discovery of imminent impactor is deemed to increase in the future,
especially because both the Flyeye telescope and NEOMIR are particularly
designed for the discovery of these special objects, with the additional goal of increasing the warning time. 
\begin{table*}
	\centering
   \caption{Near-Earth asteroids discovered before impact with Earth as of 1
   January 2025. Entries are ordered by the impacting date, and the size is estimated
   through the value of the absolute magnitude. The discovery Minor Planet
   Electronic Circular (MPEC) issued by the MPC is given as a
   reference for each asteroid.}
	\label{tab:small_impact}
        \setlength\tabcolsep{2pt} % default value: 6pt
	\begin{tabular}{lcccllcc} 
		\hline
	Designation & Meteorites  & Impact date & UTC & Discoverer & Survey & Code & Size (m)\\
		\hline
2008 TC$_3$$^{a}$  &  Y  & 2008-10-07   & 02:46 & Richard Kowalski    & Mount Lemmon Survey & G96 & 3-4     \\
2014 AA$^{b}$   &  N  & 2014-01-02   & 02:33 ($\pm 1$ h) & Richard Kowalski    & Mount Lemmon Survey & G96 & 3     \\
2018 LA$^{c}$   &  Y  & 2018-06-02   & 16:44 & Richard Kowalski    & Mount Lemmon Survey & G96 & 3-4   \\
2019 MO$^{d}$   &  N  & 2018-06-22   & 21:25 &       ---           & ATLAS-MLO           & T08 & 4-6   \\
2022 EB$_5$$^{e}$  &  N  & 2022-03-11   & 21:22 & Krisztián Sárneczky & Konkoly Observatory & K88 & 2     \\
2022 WJ$_1$$^{f}$  &  N  & 2022-11-19   & 08:27 & David Rankin        & Mount Lemmon Survey & G96 & 1     \\
2023 CX$_1$$^{g}$  &  Y  & 2023-02-13   & 02:59 & Krisztián Sárneczky & Konkoly Observatory & K88 & 1     \\
2024 BX$_1$$^{h}$  &  Y  & 2024-01-21   & 00:32 & Krisztián Sárneczky & Konkoly Observatory & K88 & 1     \\
2024 RW$_1$$^{i}$  &  N  & 2024-09-04   & 16:39 & Jacqueline Fazekas  & Mount Lemmon Survey & G96 & 1.5     \\
2024 UQ$^{j}$   &  N  & 2024-10-22   & 10:45 &       ---           & ATLAS-HKO, Haleakala& T05 & 1     \\
2024 XA$_1$$^{k}$  &  ?  & 2024-12-03   & 16:15 &  V. F. Carvajal     & Kitt Peak-Bok       & V00 & 1 \\
		\hline
  \\
  \multicolumn{8}{l}{$^{a}$ MPEC 2008-T72, \url{https://minorplanetcenter.net/mpec/K08/K08T72.html}}\\
  \multicolumn{8}{l}{$^{b}$ MPEC 2014-A02, \url{https://minorplanetcenter.net/mpec/K14/K14A02.html}}\\
  \multicolumn{8}{l}{$^{c}$ MPEC 2018-L04, \url{https://minorplanetcenter.net/mpec/K18/K18L04.html}}\\
  \multicolumn{8}{l}{$^{d}$ MPEC 2019-M72, \url{https://minorplanetcenter.net/mpec/K19/K19M72.html}}\\
  \multicolumn{8}{l}{$^{e}$ MPEC 2022-E178, \url{https://minorplanetcenter.net/mpec/K22/K22EH8.html}}\\
  \multicolumn{8}{l}{$^{f}$ MPEC 2022-W69, \url{https://minorplanetcenter.net/mpec/K22/K22W69.html}}\\
  \multicolumn{8}{l}{$^{g}$ MPEC 2023-C103, \url{https://minorplanetcenter.net/mpec/K23/K23CA3.html}}\\
  \multicolumn{8}{l}{$^{h}$ MPEC 2024-B76, \url{https://minorplanetcenter.net/mpec/K24/K24B76.html}}\\
  \multicolumn{8}{l}{$^{i}$ MPEC 2024-R68, \url{https://minorplanetcenter.net/mpec/K24/K24R68.html}}\\
  \multicolumn{8}{l}{$^{j}$ MPEC 2024-U49, \url{https://minorplanetcenter.net/mpec/K24/K24U49.html}}\\
  \multicolumn{8}{l}{$^{k}$ MPEC 2024-X68, \url{https://minorplanetcenter.net/mpec/K24/K24X68.html}}\\
	\end{tabular}
\end{table*}

Along with surveys, the currently operational monitoring systems for imminent impactors $-$
such as the Meerkat Asteroid Guard \citep{gianotto-etal_2023} by ESA, Scout\footnote{\url{https://cneos.jpl.nasa.gov/scout/}}
\citep{farnocchia-etal_2015c} by NASA, and
NEOScan\footnote{\url{https://newton.spacedys.com/neodys/NEOScan/}}
\citep{spoto-etal_2018, delvigna-etal_2021} by NEODyS $-$ have also been proven
to be efficient in identifying unconfirmed objects that could impact Earth
shortly after discovery. These services are extremely important in encouraging
the follow-up of these objects so that their orbit and possible impact location
can be better determined. 

The discovery of asteroids before impacting Earth is not only relevant for
planetary defence purposes, but it offers a wider range of possibilities for
studying their origin, if compared to fireball detections by satellite sensors
\citep{devillepoix2019, Pena-Asensio2022}. In fact, small NEAs of few meters in diameter
generally produce a bright fireball during the atmospheric entry, while
undergoing ablation and possibly multiple fragmentations. 
Some fragments may survive the airburst and experience the dark flight phase,
finally reaching the ground and producing a meteorites fall.
The area on the ground where meteorites fall is also called \textit{strewn
field}, which can be searched for in order to recover a part of the fragments.
The recovery of meteorites allows to perform accurate laboratory studies, which
give a better characterisation of the parent body. This was the case for the
past impactors 2008~TC$_3$ \citep{jenniskens-etal_2009, shaddad-etal_2010},
2018~LA \citep{jenniskens-etal_2021}, 2023~CX$_1$ \citep{bischoff-etal_2023},
and 2024~BX$_1$ \citep{bischoff-etal_2024}, while possible strewn fields have
been computed for 2022~WJ$_1$ \citep{carbognani-etal_2025, kareta-etal_2024}
although no meteorites have been recovered yet.
In addition, the knowledge of an accurate heliocentric orbit of the NEA parent
body and the physical properties of the associated meteorite may permit to
reconstruct the dynamic history of the NEA, and possibly identify progenitors
\citep{carbognani-fenucci_2023} or source regions \citep[see
e.g.][]{broz-etal_2024b, broz-etal_2024, marrset-etal_2024}.

This paper is focused on asteroid 2024~XA$_{1}$, a meter-sized object which
impacted Earth on 3 December 2024 at 16:15 UTC over a remote area of the Sakha
Republic (Russia) and was discovered only ten hours before impact. The impact
alert was first announced by Meerkat, the imminent impactor warning service
operated at the NEO Coordination
Centre\footnote{\url{https://neo.ssa.esa.int/}} (NEOCC) by ESA. At the epoch of
the alert, the asteroid was still on the NEO Confirmation
Page\footnote{\url{https://minorplanetcenter.net/iau/NEO/toconfirm_tabular.html}}
(NEOCP) of the Minor Planet
Center\footnote{\url{https://minorplanetcenter.net/}} (MPC). As soon as a
sufficient number of observations was available, a more accurate orbit and
impact corridor were computed with the ESA Aegis Orbit Determination and Impact
monitoring system \citep{fenucci-etal_2024b}. The impact point at 100 km
altitude was used to compute the possible strewn fields even prior to the
impact epoch, and it was performed by the Italian Istituto Nazionale di
Astrofisica (INAF). The final ab initio strewn field
\citep{carbognani-etal_2025} was determined with all the astrometric
observations available at the MPC, and it is reported here for future meteorite
search campaigns.

\section{Methods}

\subsection{The Meerkat imminent impactor warning system}
\label{ss:Meerkat}
The Meerkat Asteroid Guard \citep{gianotto-etal_2023} is an automated system which scans the NEOCP in search of unconfirmed near-Earth objects (NEOs) that could fly-by or impact Earth. 
Typically, NEOCP objects have a very short observational arc, and classical methods of orbit determination \citep[see, e.g.][]{milani-gronchi_2009} are usually not effective. Meerkat uses the systematic ranging technique \citep{farnocchia-etal_2015c, spoto-etal_2018} to compute the uncertainty region of an object. Orbits compatible with the given observations are then propagated for 30 days in the future to search for possible impacts with Earth. 

Early alerts of close fly-bys and Earth impactors are of extreme importance in prioritizing and encouraging astrometric follow-up of unconfirmed NEOs. To this purpose, Meerkat broadcasts the results to a dedicated mailing list opened only to people actively involved in the NEO field, particularly to observers and researchers who promptly need access to potential impactors information\footnote{Request can be sent directly to the address \texttt{neocc@esa.int}.}. 
To minimize the response time, Meerkat also makes direct phone calls to NEOCC astronomers so that they can quickly analyze the scenario and select the most suitable telescope from the NEOCC network\footnote{\url{https://neo.ssa.esa.int/neocc-observing-facilities}} to confirm the object, and possibly measure other properties \citep{devogele-etal_2024}.

\begin{figure*}
    \centering
    \includegraphics[width=0.8\textwidth]{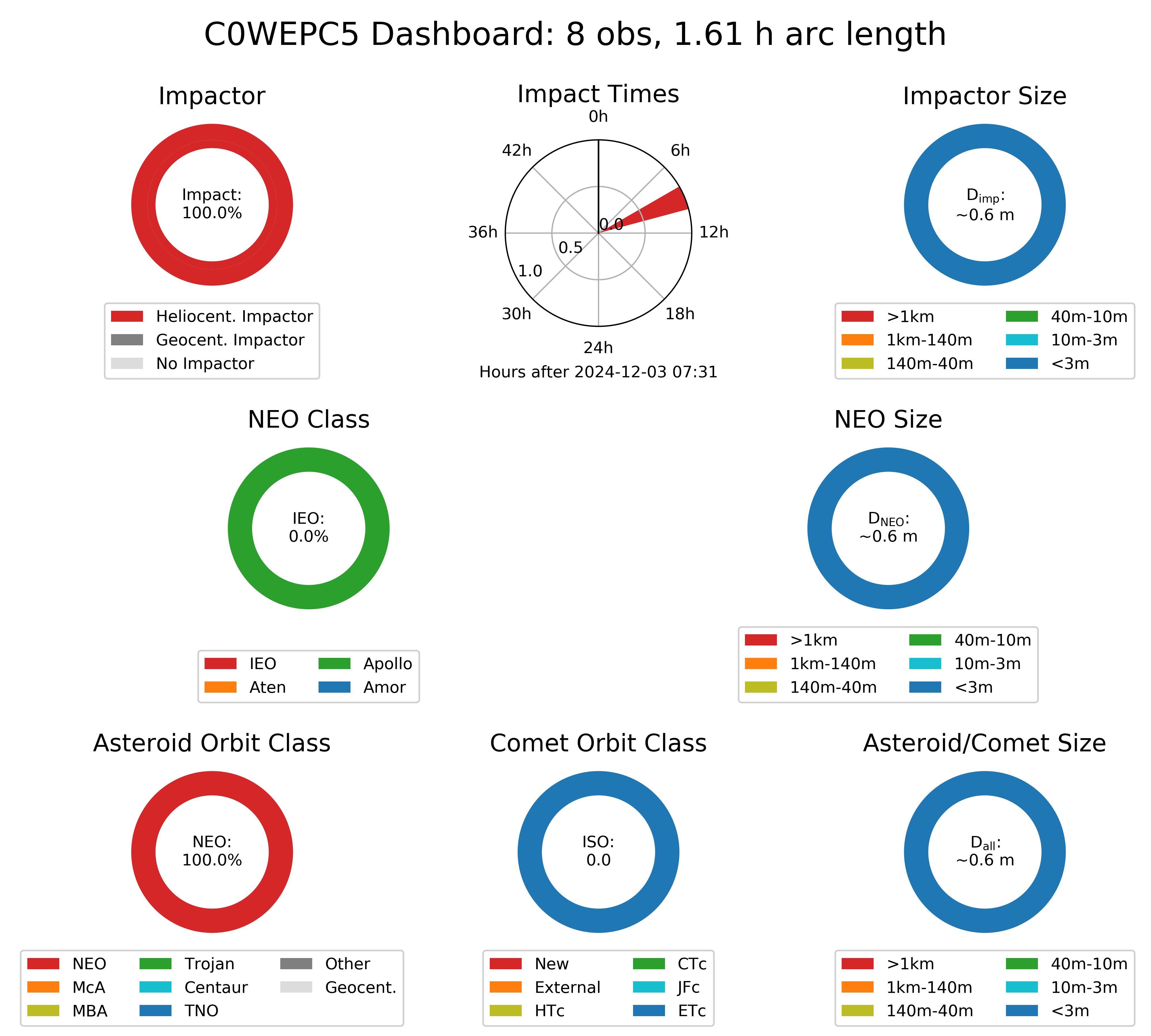}
    \caption{Typical view of the Meerkat Dashboard. This example is the one sent to subscribers in the first impact alert of 2024~XA$_1$.}
    \label{fig:meerkat_first_alarm_dashboard}
\end{figure*}

Results obtained by Meerkat are presented to users in a visual manner, to easily understand the properties of the selected objects.
A dashboard containing several pie charts give clear information about the orbit class and the impact threat (see Fig.~\ref{fig:meerkat_first_alarm_dashboard} for an example). The first row of the dashboard presents an impact score, an impact time chart, and the size of the impactor computed from the absolute magnitude of the impacting orbital solutions. The second and the third rows give information about the NEO and comet classes, along with a class-dependent size estimate.
The asteroid size is computed from the absolute magnitude $H$ over the whole 95\% confidence region obtained by the systematic ranging. On the other hand, the impactor size is obtained by taking only the size of the impacting solutions within the 95\% confidence region. The conversion from absolute magnitude to diameter is performed through the formula $D = 1329 \textrm{ km} \times 10^{-H/5}/\sqrt{p_V}$ \citep{bowell-etal_1989, pravec-harris_2007}, where the albedo $p_V$ is fixed to 0.14. The distinction between asteroid size and impactor size is important because impacting solutions of the systematic ranging generally tend to have larger absolute magnitudes, and hence smaller object sizes. Without this classification, mitigation and follow-up actions might be based on too large impactor size estimates. Finally, note that for the case of 2024~XA$_1$ presented in Fig.~\ref{fig:meerkat_first_alarm_dashboard}, the two sizes were the same because the impact probability with 8 observations already reached 100\%.

\begin{figure}
    \centering
    \includegraphics[width=0.6\linewidth]{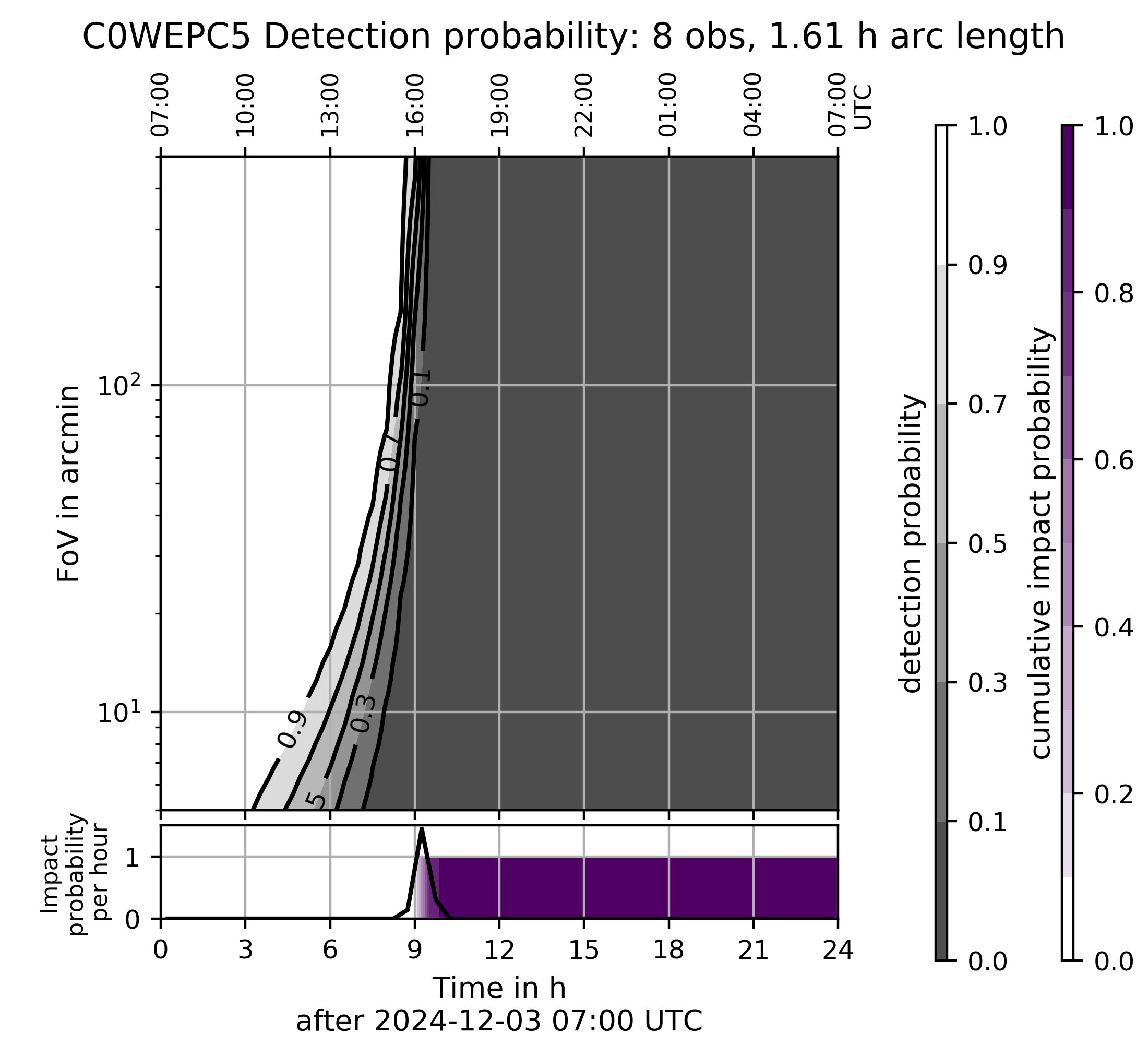}
    \caption{Station selector plot for 2024~XA$_1$ computed with 8 observations, showing the contour lines of the detection probability. The bottom panel shows the cumulative impact probability as a function of the time.}
    \label{fig:station_selector}
\end{figure}

A station selector plot shows the detection probability as contour lines,
depending on the epoch of observation and the size of the field of view. The
detection probability values are not station-specific, but geocentric.
The cumulative impact probability as a function of the time is also
shown in a panel on the bottom of the figure. Figure~\ref{fig:station_selector}
shows the plot for 2024~XA$_1$ computed with 8 observations.

\begin{figure}
    \centering
    \includegraphics[width=0.8\linewidth]{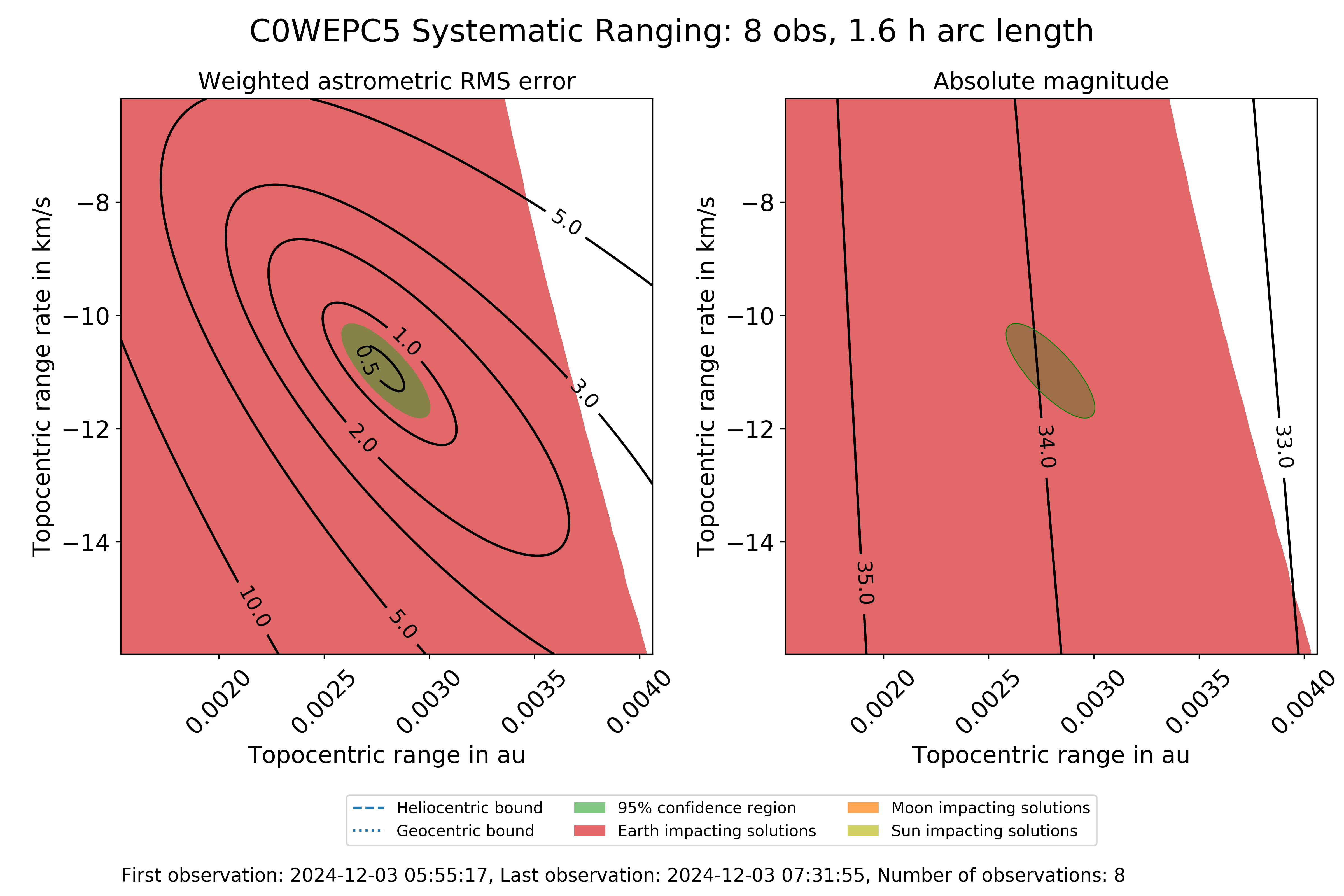}
    \caption{Systematic ranging of 2024~XA$_1$ performed with 8
    observations. The left plot shows the contour lines of the weighted
    root-mean-square (RMS) error of the astrometric residuals, in the plane of
    topocentric range $\rho$ and topocentric range rate $\dot{\rho}$. The right
    plot shows the contour lines of the absolute magnitude, in the same plane
    $(\rho, \dot{\rho})$. The 95\% confidence region of the orbit and the
    region of Earth impacting solutions are also shown in both panels, in green
    and red areas respectively. }
    \label{fig:systematic_ranging}
\end{figure}

Results of the systematic ranging are given in two plots, both in the
plane of topocentric range and topocentric range rate. The left panel shows the
contour lines of the weighted root-mean-square error of the astrometric
residuals, while the right panel shows the contour lines of the absolute
magnitude. The 95\% confidence region of the orbit and the region of Earth
impacting solutions are also shown in the plot.
Figure~\ref{fig:systematic_ranging} shows an example of the systematic ranging
plot, obtained on 2024~XA$_1$ with 8 observations. 

\begin{figure}
    \centering
    \includegraphics[width=0.8\linewidth]{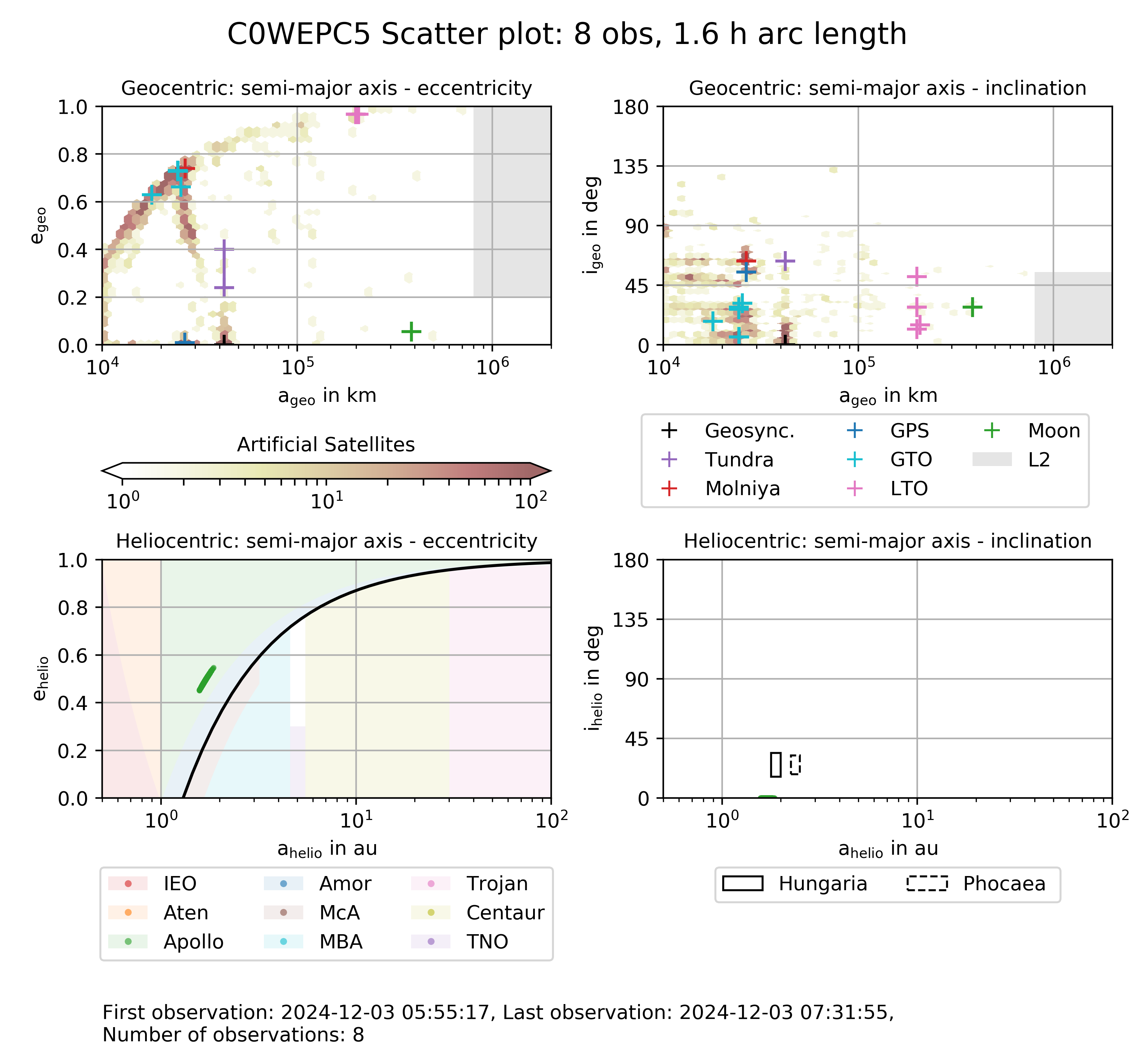}
    \caption{Scatter plots of 2024~XA$_1$ computed by Meerkat with 8
    observations. The top-left panel shows the density of artificial satellites
    in the plane of geocentric semi-major axis and geocetric eccentricity.
    Coloured crosses indicate the location of different type of artificial
    satellites, e.g. geosynchronous, global positioning system (GPS), Molniya,
    etc. The top-right panel shows the same density in the plane of geocentric
    semi-major axis and geocentric inclination. Geocentric solutions from
    systematic ranging would be plotted with black dots. The bottom-left plot
    shows the plane of semi-major axis and eccentricity of heliocentric orbits,
    divided into different color-coded orbital categories (such as Amors,
    Apollo Atens, etc). The black curve corresponds to the boundary of the NEO
    region, defined by a perihelion distance of 1.3 au. Finally, the
    bottom-right panel shows the location of the Hungaria and Phocaea regions
    in the plane of semi-major axis and inclination. The heliocentric orbital
    solutions computed for the unconfirmed object by the systematic ranging are
    showed with green dots in the plots of the second row. }
    \label{fig:scatter}
\end{figure}

Scatter plots in the planes of geocentric (heliocentric) semi-major axis $-$
eccentricity and semi-major axis $-$ inclination are useful to understand the
orbital elements of the object, and possibly make a guess on its
nature. The plots in the geocentric elements come along with the background
density of artificial satellites to easily identify if a solution given
by the systematic ranging corresponds to a region densely populated
with artificial satellites. The plots in the heliocentric elements are, on the
other hand, useful to understand the possible source region of the object.
Figure~\ref{fig:scatter} shows the scatter plots produced by Meerkat
for 2024~XA$_1$, computed with 8 observations. Note that the object does not
appear in the plots with geocentric elements, which indicate that with already
8 observations it was possible to exclude the possibility of an artificial
object orbiting Earth. It was also possible to constrain the orbital elements
to an Apollo-type orbit.

Finally, a map showing the impact positions on Earth is given along with the alert message (see Fig.~\ref{fig:meerkat_first_IC} for an example).

\subsection{Orbit determination and impact predictions}
\label{ss:Aegis}
The NEOCC operates also the Aegis Orbit Determination and Impact Monitoring system \citep{fenucci-etal_2024b}. This automated system is designed to determine the orbits of asteroids already designated by the MPC and compute the impact probabilities with Earth in the next 100 years. The orbit determination implemented in Aegis is based on a least-square method, which is solved through an iterative differential corrections algorithm. 
Initial conditions for the differential corrections can be computed through the Gauss or Laplace initial orbit determination methods \citep[see][for an overview]{milani-gronchi_2009}. 
The impact monitoring is based on the Line of Variations (LOV) method, which is
a geometrical 1-dimensional sampling of the confidence region
\citep{milani-etal_2005b}. This method relies on the fact that the confidence
region is usually stretched along a \textit{weak direction}, over which the
uncertainty is the largest. The LOV is sampled with a relatively small number
of Virtual Asteroids (VAs), which are then propagated in the future to search
for Virtual Impactors (VIs). The LOV method also allows to compute an impact
probability for a given VI. More details about these methods can be found in,
e.g., \citet{milani-etal_2002, milani-etal_2005}. When a VI with a
sufficiently large impact probability is found, the impact corridor can be
computed through a semi-analytical algorithm \citep{dimare-etal_2020}. More
details about the Aegis system can be found in \citet{fenucci-etal_2024b} and
references therein.

In the case of imminent impactors, many observations may be available thanks to the prompt follow-up, and classical methods for orbit determination and impact monitoring are more accurate than the systematic ranging implemented in Meerkat. Therefore, we developed an automated pipeline to process NEOCP objects using the Aegis software. The pipeline is triggered when an impact probability larger than 10\% is found by Meerkat, and the corresponding tracklet has more than 4 observations. Meerkat sends the tracklet to the Aegis system, and a preliminary orbit is computed using the Gauss method. This orbit is then used as a starting guess to fit a least-square orbit. After, impact monitoring based on the LOV method is performed to search for possible impacts in the next 30 days. Finally, the impact corridor at 100 km and 0 km altitudes are computed without taking into account atmospheric effects in the determination of the ground impact point. 
Since this pipeline is dedicated to unconfirmed objects, the results of the computations are not published on the NEOCC web portal, but they are sent to NEOCC staff by email.     

\subsection{Ab initio strewn field computation}

In \citet{carbognani-etal_2025}, the authors computed the strewn field of three
past impacted NEAs, namely 2008~TC$_3$, 2023~CX$_1$, and 2024~BX$_1$ (see
Table~\ref{tab:small_impact}), with uncertainties of about one kilometer
across the strewn field, which is a distance which could be feasible
to cover by walking. Here we use the same fall model, which we briefly
summarize here. Starting from the entry point at 100 km above the Earth
surface, the motion of the meteoroid entering the atmosphere is modeled by a
force model which includes the Earth gravity and the drag force. The fall model
considers only one main fragmentation, which happens when the aerodynamic
pressure exceeds the mechanical strength $S$ of the meteoroid. During
fragmentation, we assume that the meteoroid disrupts into a set of smaller
fragments with reasonable a-priori mass values. All the fragments are assigned
the same initial velocity, without the addition of a lateral component. The
points on the Earth surface where these fragments fall define the strewn field
and, because the fall zone is computed only with data prior to impact, this
computation is called \textit{ab initio strewn field}.

The wind profile used to model the drag force during the fall of 2024~XA$_1$ was determined by Meteo
Expert\footnote{\url{https://www.meteo.expert/}}, a private company which
provides weather predictions computed with internally developed models. Meteo Expert provides paid services, and anyone can request data, but in this case, it was a scientific collaboration in agreement with the structure and without charges. All the
available data from the surface to the troposphere and remote sensing measurements are
integrated in the atmospheric prediction model. This way, the wind profile can
be obtained near the dark flight starting point, rounding the time to the
integer hours closest to the impacting epoch.

Some assumptions are also needed in order to use the described fall model.  The
strength $S$ of the asteroid is generally unknown since it depends on its
composition and internal structure.  However, strength determines the
height at which the fragmentation takes place, which could 
significantly influence the resulting strewn field.  Therefore, we performed the
computations with different strength values, namely $S = 0.5, 1, 5$ MPa. These
values had already given good results in the analysis of the strewn
fields of 2024~BX$_1$, 2023~CX$_1$ and 2008~TC$_3$ in
\citet{carbognani-etal_2025} and are in the strength range estimated from the
observed fireballs \citep{Borovicka2008}, thus this approach is effective in
determining at least one ab initio strewn field close to the real one.

Another critical parameter is the mass range of the fragments, which is also
unknown. In our computations, we assumed a-priori fragment masses of 1, 0.3,
0.2, 0.1, 0.05, 0.02, 0.005, and 0.001 kg, which are typical values for
meteorites originating from small meter-sized asteroids as 2024~XA$_1$. It must
be pointed out that, although it is reasonable to search for meteorites in the
mass range 1 g - 1 kg, there is no guarantee that such fragments may be
recovered. In fact, the fall event could have various outcomes, including the
two rare extreme cases of complete disintegration $-$ typical of cometary
bodies $-$, or no fragmentation at all \citep{Borovicka2008, Kenkmann2009}.

Finally, the model assumes that the asteroid undergoes only one fragmentation
event during atmospheric entry. However, results showed in
\citep{carbognani-etal_2025} indicated that the presence of multiple
fragmentations do not significantly affect the position of the final strewn
field, provided that the impacting object underwent at least a main
fragmentation.

\section{Results}
\subsection{The first Meerkat impact warning}
\begin{figure}
    \centering
    \includegraphics[width=0.9\textwidth]{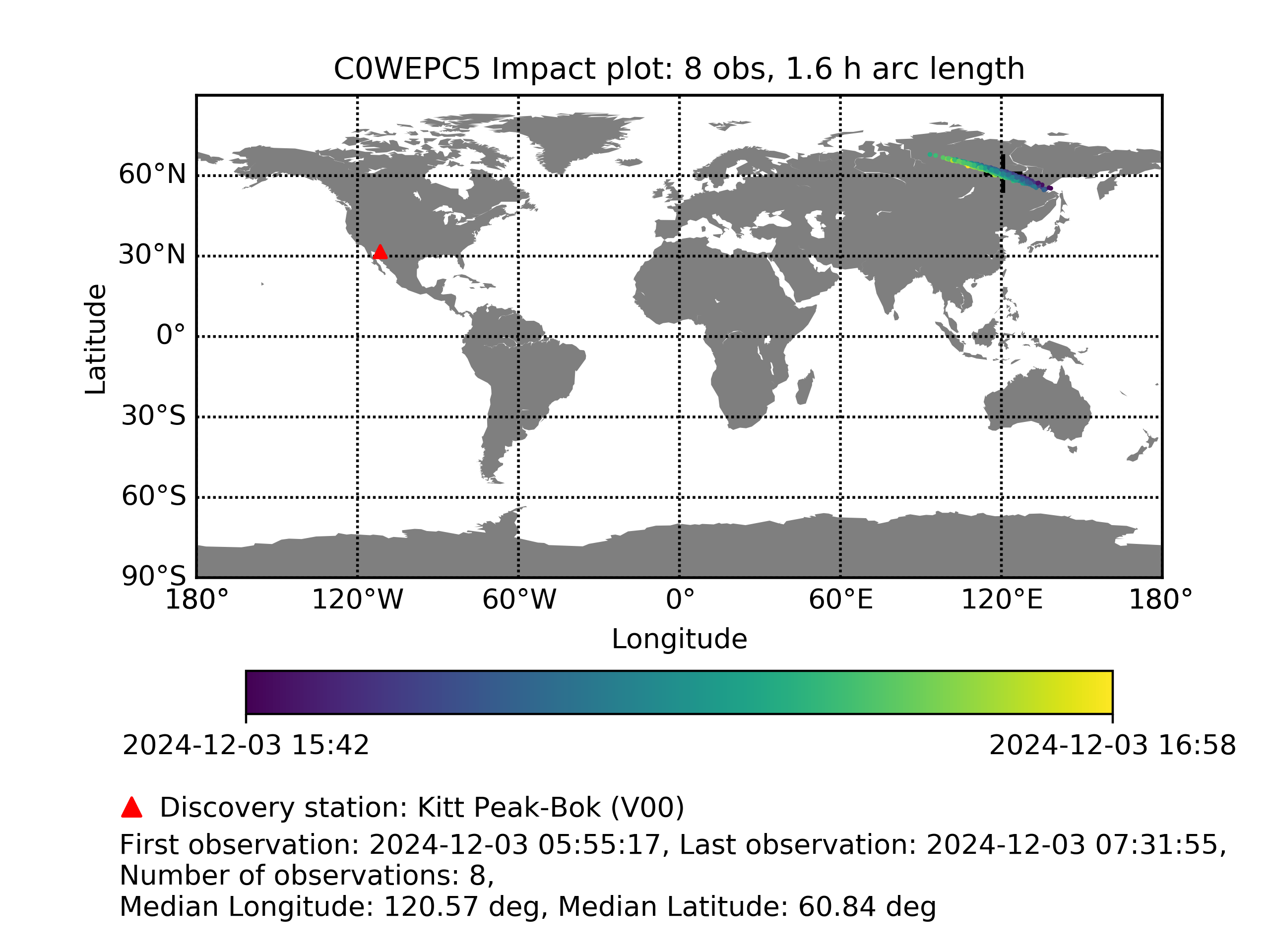}
    \caption{Impact location computed by Meerkat and sent to subscribers in the first impact alert.}
    \label{fig:meerkat_first_IC}
\end{figure}

Asteroid 2024~XA$_1$ was discovered at 05:54 UTC on 3 December 2024 by V. F. Carvajal using the Bok telescope in Kitt Peak, Arizona (MPC code V00). A first tracklet containing only 4 observations upon discovery was posted on the NEOCP, and the object was provisionally designated as C0WEPC5. With this set of observations, all the imminent impactor warning systems reported a very small chance of impact with Earth, smaller than 1\%. 
About an hour after discovery, a follow-up set of 4 observations was taken by the Steward Observatory in Mount Lemmon (MPC observatory code I52) and announced on the NEOCP at 07:46 UTC. At 07:50 UTC, Meerkat sent an impact alert on the mailing list, announcing that C0WEPC5 had a 100\% impact probability with Earth between 15:38 UTC and 17:06 UTC. The estimated size of the object was 0.6 m.  Figure~\ref{fig:meerkat_first_alarm_dashboard} shows the Meerkat dashboard sent in the first alert. The first impact corridor computed by Meerkat (see Fig.~\ref{fig:meerkat_first_IC}, bottom panel) indicated a median impact location over the Sakha Republic (Russia) at coordinates Lat. 60.97$^\circ$ N Long. 119.99$^\circ$ E, with large uncertainties due to the small number of observations. Impact alerts confirming Meerkat's results were announced a few minutes later by NASA Scout and NEOScan by NEODyS. 
The Aegis pipeline described in Sec.~\ref{ss:Aegis} was triggered after Meerkat's computations, and it confirmed the impact with large uncertainties in the entry area.

\subsection{Astrometric follow-up}
During the first alert, Meerkat also successfully notified NEOCC astronomers by phone calls, and procedures to follow-up the object to obtain more observations started. At the time of the discovery trigger, the object was already too low to observe from any of our facilities in the Chilean Andes. Our initial attempts to obtain follow-up were therefore directed to telescopes in the US South-West, but they were hampered by poor weather. We were, however, able to successfully schedule an observing block with one of the 0.35 m telescopes of the Las Cumbres Observatory in Haleakala, Hawaii (MPC code T03). After Hawaii, the best observability conditions crossed the empty space of the Northern Pacific Ocean, leaving few observational opportunities until becoming observable from Eastern Asia or Oceania. However, our network telescopes in Japan and Australia were also affected by poor weather, and therefore, we had to wait until about one hour before shadow entrance to successfully obtain images of the object with the 2.0 m Himalaya Chandra Telescope in India (MPC code N50). Unfortunately, the object had already been impacted by the time it got dark on any European telescope in our network, thus ending our observational coverage.

During the same hours, we were in contact with other independent observers worldwide, who had easier telescope access and were able to extract valuable astrometry from their datasets. The general follow-up observations from code G37 and F65, discussed below, were provided by external collaborators but actually measured and reported in near real-time by our team.

Before impact, eight other updates were posted on the NEOCP, and impact alerts continued to be triggered. The availability of more astrometry allowed the uncertainties in both the time of impact and the location to shrink significantly. The object entered the Earth's shadow 20 minutes before impact, and it was impossible to observe it anymore. A total of 40 observations were available before impact from 7 different observatories (MPC codes I41, V00, I52, H01, G37, F65, C94).

\subsection{Impact event, heliocentric orbit, and impact location}
\begin{figure*}
    \centering
    \includegraphics[width=0.9\textwidth]{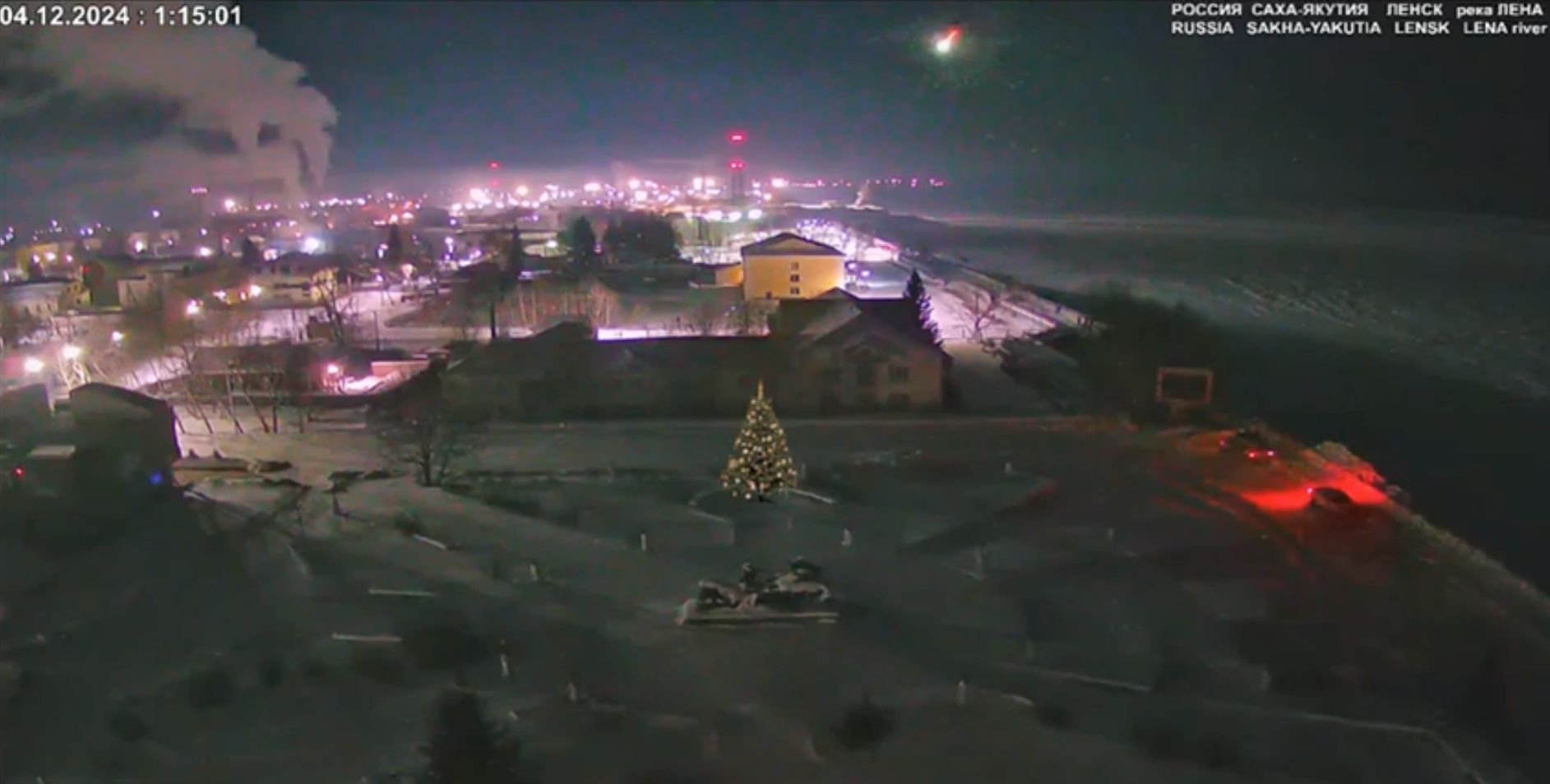}
    \caption{Frame from a live webcam in Lensk (Sakha Republic, Russia) capturing the atmospheric entry of 2024~XA$_1$. The fireball can be seen at the top of the frame.}
    \label{fig:Lensk_Camera}
\end{figure*}

The last Meerkat alert before impact, sent at 14:47 UTC, was obtained with 40 observations and predicted the impact at  16:14:53 $\pm$ 9 s UTC, with a few kilometers of uncertainty in the impact position.
To follow the impact event and check the accuracy of the predictions, we identified a live webcam in Lensk (Sakha Republic, Russia) with a field of view pointing at the predicted entry direction. The fireball appeared in the webcam live streaming at the predicted time, and the entire video is available on YouTube\footnote{\url{https://www.youtube.com/watch?v=iKlGwZH9i0Y&t}}. Figure~\ref{fig:Lensk_Camera} shows a frame imaging the fireball.

\begin{figure*}
    \centering
    \includegraphics[width=0.9\textwidth]{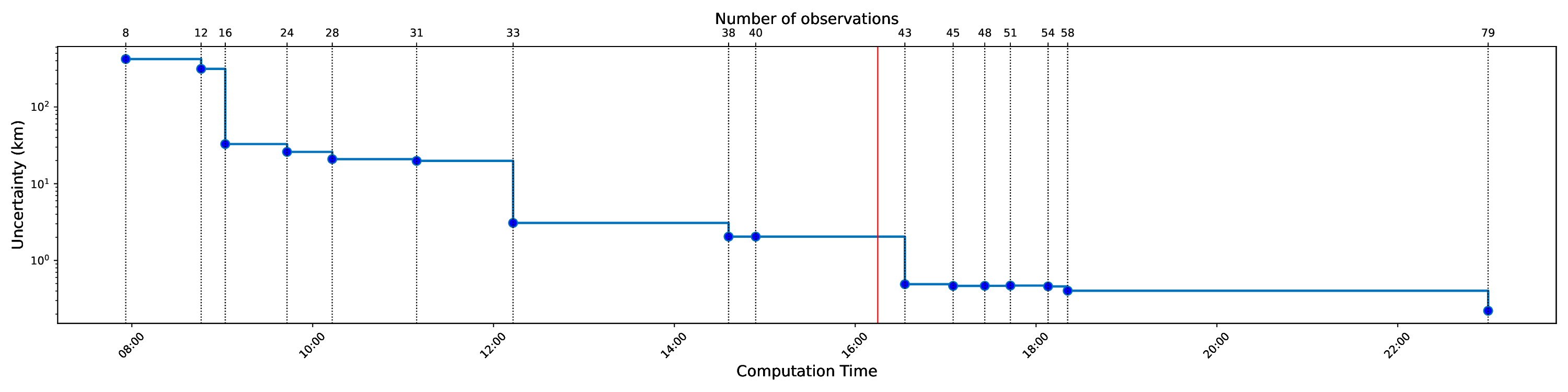}
    \caption{Uncertainty in the impact location as new observations were announced on the NEOCP. The lower x-axis corresponds to the computation time, while the upper x-axis indicates the corresponding number of observations used for the computation. The y-axis reports the 1-$\sigma$ uncertainty of the impact location at 100 km altitude. The vertical red line denotes the time of impact. The solution with 79 observations was computed on 17 December, but it was reported on the day after the fall to keep the readability of the plot.}
    \label{fig:IC_evol}
\end{figure*}

Additional astrometry was posted on the NEOCP even after impact, and the Aegis pipeline continued to compute the impact corridor with the available data. Figure~\ref{fig:IC_evol} shows the uncertainty of the impact location as a function of the observations available on the NEOCP and the time of computation. At the beginning, the uncertainty was of the order of 400 km, and it fell below 10 km when 33 observations were posted on the NEOCP. With the observations available at the time of impact, the uncertainty shrinks to about 1 km. The object was finally designated as 2024~XA$_1$ and announced in MPEC 2024-X68\footnote{\url{https://minorplanetcenter.net/mpec/K24/K24X68.html}}, which was issued at 18:05 UTC of 3 December.

On 17 December, we downloaded all the observations available at the MPC from the MPC Explorer\footnote{\url{https://data.minorplanetcenter.net/explorer/}} service, using the new Astrometry Data Exchange Standard (ADES) format. We then accurately recomputed the heliocentric orbit of 2024~XA$_1$, using all the 79 available astrometric uncertainties reported by the observers. Table~\ref{tab:helio_orbit} reports the orbital elements on 1 October 2024\footnote{Note that in this case, the orbital elements at the middle epoch of observations would be significantly perturbed already by Earth's gravitational attraction.}. The final impact point at 100 km altitude had a 1-$\sigma$ uncertainty of 220 m, slightly larger than the recent impactors 2023~CX$_1$ and 2024~BX$_1$.

{\renewcommand{\arraystretch}{1.2} % Change row height for this table only
\begin{table}[!ht]
    \centering
    \caption{Keplerian orbital elements of 2024~XA$_1$ on 1 October 2024. Errors refer to the 1-$\sigma$ formal uncertainties.}
    \begin{tabular}{lr}
         \hline
         \hline
      Parameter                      & Value  \\
         \hline
      Epoch                          & 2024-10-01 00:00:00 UTC  \\ %57935.195253400 TDT 
      Semi-major axis, $a$ (au)                &  $  1.68911 \pm  5.1\times 10^{-5}$ \\ 
      Eccentricity, $e$                        &  $  0.49518 \pm  1.7\times 10^{-5}$ \\ 
      Inclination, $i$ (deg)                   &  $  0.10824 \pm  8.4\times 10^{-6}$ \\ 
      Longitude of node, $\Omega$ (deg)        &  $ 72.26181 \pm  6.7\times 10^{-5}$ \\ 
      Argument of perihelion, $\omega$ (deg)   &  $ 53.08039 \pm  9.0\times 10^{-5}$ \\ 
      Mean anomaly, $\ell$  (deg)              &  $313.99293 \pm  9.1\times 10^{-4}$ \\ 
         \hline
    \end{tabular}
    \label{tab:helio_orbit}
\end{table}
}

\begin{table}[!ht]
    \centering
    \caption{Them mean trajectory parameters at 100 km altitude (WGS84 ellipsoid) from Earth surface of 2024~XA$_1$ given by heliocentric orbit with their 1-$\sigma$ uncertainties. The speed is relative to the ground. The inclination is the angle that the speed vector forms with the ground. The azimuth, counted from north to east, gives the fireball's incoming direction. %The JPL solution n.3 with 54 observations (Farnocchia, NEOCP mailing list) is similar to ours: Lat.=60.6287$^\circ$ N, Long.=119.0741$^\circ$ E, Speed=15,524 km/s, Azimuth =214.96$^\circ$ and Inclination= 50.61$^\circ$.
    }
    \begin{tabular}{ccc}
    \hline
       Object          & 2024~XA$_1$  \\
    \hline
       Time (UTC)                     & 2024-12-03 16:14:52.87 $\pm$ 0.070 s  \\
       Latitude (deg)                 & $60.6285   \pm 0.0007$    \\
       East Longitude (deg)           & $119.0713  \pm 0.002$  \\
       Speed (km s$^{-1}$)            & $15.5241    \pm 0.0001$    \\
       Inclination (deg)              & $50.6116  \pm 0.001$     \\
       Azimuth (deg)                  & $214.9572  \pm 0.003$  \\
       \hline
    \end{tabular}
    \label{tab:parameters}
\end{table}

\subsection{Live strewn field and possible final strewn fields}
\begin{figure}
    \centering
    \includegraphics[width=0.9\textwidth]{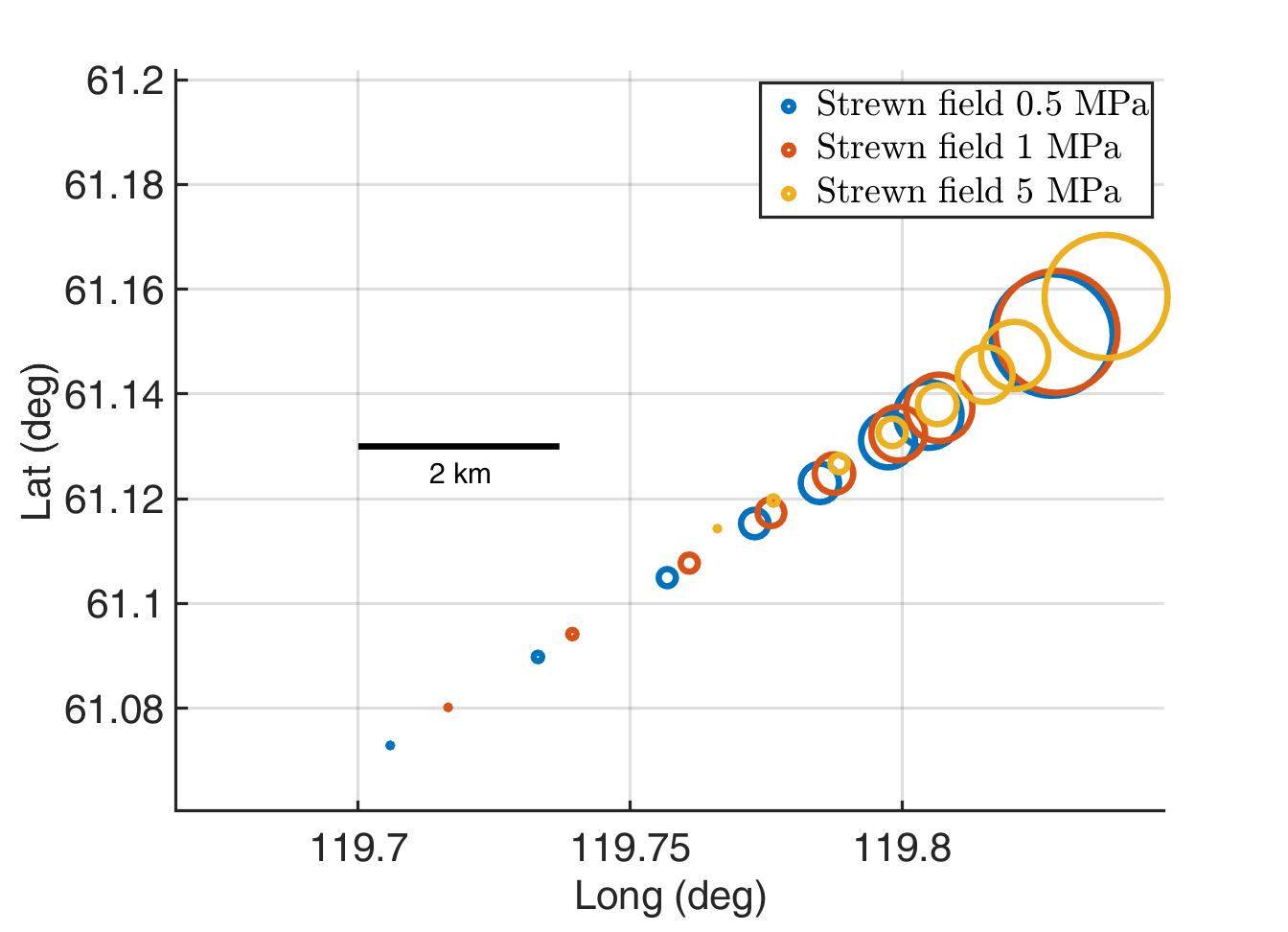}
    \caption{The possible strewn fields of 2024~XA$_1$ with a mean strength of 0.5, 1 and 5 MPa. The circles are proportional to the mass of the possible meteorite. There can be no certainty that fragments of 2024 XA1 have reached the ground, but this area is where it is most likely to find something. The uncertainty about the location of possible fragments is on the order of 1 km.}
    \label{fig:2024XA1_strewn_field}
\end{figure}

The first preliminary computations of the \textit{ab initio} strewn field began
around 8 UTC on 3 Dec 2024, when it was clear that 2024~XA$_1$ would hit Earth.
The impact site was not yet well defined (see Fig.~\ref{fig:IC_evol}), but
taking as reference the median geographic coordinates of the possible entry
points into the atmosphere at 100 km altitude, a provisional atmospheric
profile was acquired from radio-sounding data taken at 00 UTC on 3 December
from the Olekminsk station (Lat. $60.37^\circ$ N, Long. $120.42^\circ$
E)\footnote{\url{https://weather.uwyo.edu/upperair/europe.html}}. This made it
possible to make a first estimate of the average geographic coordinates of the
beginning of the dark flight of the 1 kg fragment as a function of the possible
asteroid strength $S$ of 0.5, 1 and 5 MPa, which worked well in the cases of
the 2024~BX$_1$, 2023~CX$_1$, and 2008~TC$_3$. With this provisional data,
Meteo Expert computed the first real atmospheric profile for 16 UTC,
which made it possible to have the first reliable estimate of the
possible strewn fields. This procedure was repeated a couple of times
because the orbit was constantly updating, and at 15 UTC, about two
hours before the fall, with 40 astrometric observations, we already had a good
estimate of the area of the possible strewn fields until arriving at the
definitive result, using the full astrometric set of measures, that we show in
Fig.~\ref{fig:2024XA1_strewn_field}. Compared to the final version, the
preliminary strewn field is approximately 1.6 km to the east.

According to the model results, with an average strength of $S=0.5$ MPa, the main fragmentation occurred at 40.6 km altitude, a value that drops to 35.6 km for 1 MPa and 23.7 km for $S=5$ MPa. From the live webcam that captured the fireball, we know there are at least two flares, one of which was very intense towards the end of the fireball phase, so the asteroid fragmented during the fall, see Fig.~\ref{fig:Lensk_Camera}. For all three possible strength cases, the geographic coordinates of the starting point of the dark flight of the possible fragment with a final mass of 1 kg are Lat. $61.1^\circ$ E, Long. $119.7^\circ$ E. Due to the high inclination of the trajectory, the three possible strewn fields largely overlap, which is an advantage for ground research, see Fig.~\ref{fig:2024XA1_strewn_field}. The potential 0.5 MPa strewn field has a ground length of 10 km, a value that drops to 6 km for the 5 MPa one because the dispersion of the fragments began at a lower height. In either case, an eventual meteorite search campaign would occur within walking distance. The strewn fields are located about 13 km ahead of the theoretical impact point without considering the atmosphere, 37 km northeast of Kiliyer village, 89 km north of the city of Olekminsk, and 950 km east of Tunguska's epicentre (see Table~\ref{tab:strewn_fields}). From Google Maps, the area appears to be covered by Taiga vegetation, so searching on the ground will not be easy.

\begin{table}[!ht]
    \centering
    \caption{The possible nominal strewn fields coordinates. The column Mass shows the mass of the hypothetical meteorite, while the columns with $S$ show the possible mean asteroid's strength with lat N and long E coordinates in degrees of the possible meteorite. Based on the results obtained on 2024~BX$_1$, 2023~CX$_1$ and 2008~TC$_3$, the uncertainty on the position of the meteorites is of the order of 1 km \citep{carbognani-etal_2025}.}
    \begin{tabular}{lrrr}
    \hline
       Mass (g)   & $S=0.5$ MPa & $S=1$ MPa & $S=5$ MPa\\
    \hline
       1000       & 61.151     & 61.152 & 61.159 \\
                  & 119.828     & 119.828 & 119.837 \\
       300        & 61.136     & 61.137 & 61.147 \\
                  & 119.805     & 119.807 & 119.821 \\
       200        & 61.131     & 61.132 & 61.144 \\     
                  & 119.797     & 119.799 & 119.815 \\
       100        & 61.123     & 61.125 & 61.138 \\ 
                  & 119.785     & 119.787 & 119.806 \\
       50         & 61.115     & 61.117 & 61.133 \\ 
                  & 119.773     & 119.776 & 119.798 \\
       20         & 61.105     & 61.108 & 61.127 \\ 
                  & 119.757     & 119.761 & 119.788 \\
       5          & 61.090     & 61.094 & 61.120 \\ 		 
                  & 119.733     & 119.739 & 119.776 \\
       1          & 61.073     & 61.080 & 61.114 \\
                  & 119.706     & 119.716 & 119.766 \\
       \hline
    \end{tabular}
    \label{tab:strewn_fields}
\end{table}

\section{Conclusions}
In this paper, we have reconstructed the last life hours of asteroid 2024~XA$_{1}$, discovered at 05:54 UTC on 3 December 2024, from the Bok telescope at Kitt Peak Observatory, Arizona. At the time of discovery, the asteroid was of magnitude +20 and was about 420\,000 km from Earth. The asteroid, provisionally designated C0WEPC5, has been included in the NEOCP on the MPC and confirmed by other observatories. At the discovery epoch, the probability of impact estimated by the rapid warning systems (Meerkat by ESA, Scout by NASA, and NEOScan by NEODyS) was of the order of 1\%, while it suddenly increased to 100\% as soon as the first follow-up observations were available, as notified by Meerkat at 07:50 UTC. Impact alerts confirming Meerkat's results were announced a few minutes later by NASA Scout and NEOScan by NEODyS. Initially, the impact point at 100 km altitude had a 1-$\sigma$ uncertainty of the order of 400 km and shrunk to about 1 km with the observations available at the time of impact, about 16:15 UTC. With all the astrometric observations available from MPC, the final uncertainty is 220 m with the trajectory starting parameters.\\

We tried to answer the question of where the meteorites ended up. Near the area of the fall, there are no all-sky camera networks to triangulate the fireballs, and the fall was filmed by chance by a live webcam and by occasional witnesses with smartphones, so we compute a strewn field \textit{ab initio}. Assuming an average strength of 0.5, 1 and 5 MPa for the asteroid and considering the local atmospheric profile computed ad hoc, we estimated the position of the possible theoretical strewn fields (see Table~\ref{tab:strewn_fields}). In this case, the asteroid arrived with a fairly high inclination with respect to the Earth's surface (about $51^\circ$), so the possible strewn fields overlap, facilitating the search. Since the impact occurred in a remote area of the Sakha Republic (Russia) with difficult environmental conditions, we do not expect an immediate search. Still, we hope some fragments can be recovered in future search campaigns.

\section*{Acknowledgments}
We thank the two anonymous referees for their time and constructive feedback, which have helped improve the quality of this paper.

%-------------------------------------------------------------------
\bibliographystyle{cas-model2-names}
\bibliography{holybib.bib}{} 

\end{document}